\documentclass[final,5p,twocolumn]{elsarticle}

\usepackage{graphicx}          
\usepackage[dvips]{epsfig}    

\usepackage{amssymb}
\usepackage{amsthm}

\usepackage{amsmath}          

\usepackage{algorithmic}

\journal{Mechatronics}

\begin{document}

\begin{frontmatter}

\title{On break-away forces in actuated motion systems with nonlinear friction} 

\author{Michael Ruderman}
\ead{michael.ruderman@uia.no}


\address{Faculty of Engineering and Science,
University of Agder, Grimstad, 4879-Norway}

\begin{abstract}                          
The phenomenon of so-called break-away forces, as maximal
actuation forces at which a sticking system begins to slide and
thus passes over to a steady (macro) motion, is well known from
engineering practice but still less understood in its cause-effect
relationship. This note analyzes the break-away behavior of
systems with nonlinear friction, which is analytically
well-described by combining the Coulomb friction law with
rate-independent presliding transitions and, when necessary,
Stribeck effect of the velocity-weakening steady-state curve. The
break-away conditions are harmonized with analytic form of the
system description and shown to be in accord with a relationship
between the varying break-away force and actuation force rate --
well known from the experiments reported in several independently
published works.
\end{abstract}

\begin{keyword}
Friction \sep break-away force \sep nonlinearity \sep presliding
\sep modeling \sep stiction force \sep hysteresis
\end{keyword}

\end{frontmatter}

\newtheorem{thm}{Theorem}
\newtheorem{lem}[thm]{Lemma}
\newtheorem{clr}{Corollary}
\newdefinition{rmk}{Remark}
\newproof{pf}{Proof}

\section{Introduction}
\label{sec:1}

The break-away force and related break-away conditions belong to
significant and well-known, but still not fully studied, aspects
of nonlinear dynamic friction in the actuated motion systems. The
break-away as phenomenon can be seen as a brief yet non-discrete
transition between the presliding and gross sliding when an idle
system with friction is subject to a continuously increasing
actuation (input) force. Due to the lack of measurement access and
complexity of nonlinear friction transitions the break-away
instant and force are particularly challenging for accurate
detection and description in a closed analytic form. Some
researchers even noted that "quantitative prediction of the
break-away friction level seems not yet possible"
\cite{altpeter1999friction}.

The first detailed studies of presliding frictional
characteristics and transitions into gross sliding may be credited
to the works of Dahl, e.g. \cite{Dahl68,dahl1976solid}. Later, in
the well-celebrated survey on friction modeling and control
\cite{armstrong1994survey} the authors also addressed the
break-away friction while noting that the break-away is not
instantaneous and the corresponding modeling should account for
translational distance. In further works on dynamic friction
modeling
\cite{deWit1995new,iwasaki1999disturbance,swevers2000integrated}
the authors have paid attention to, and extracted from the
numerical simulations, a dependency of the break-away force on the
actuation force rate. In favor to that quite similar relationships
have been demonstrated in various experimental setups in
\cite{johannes1973role,iwasaki1999disturbance,lampaert2004experimental}.
Further accurate measurements of the presliding friction
transitions, continuous sliding, static friction, and dynamic
friction effects can be found in
\cite{heslot1994creep,socoliuc2004transition,lee2007static,harnoy2008modeling}.

Despite the break-away phenomenon is well known from the
engineering practice and has been addressed, or at least
mentioned, is several studies on the kinetic friction, its
modeling and control, less work has been dedicated to formulate
the straightforward analytic conditions and derive the expressions
for break-away, which would be in line with the corresponding
system modeling. It seems that an explicit analysis and math
notation of break-away states has been solely provided in
\cite{armstrong1993stick}, while the break-away force has been
rather considered as a function of dwell time, and the given
deviations seem less suitable for a direct practical use.

With this note we address the relationship between the break-away
friction force and actuation force rate in a possibly simple and,
at the same time, coherent way based on the established modeling
assumptions and results published in several independent works.
The following analysis and presentation should contribute to
better understanding the frictional break-away behavior and help
in predicting and controlling the actuated presliding transitions.

\section{Sliding and presliding friction}
\label{sec:2}

The tangential friction force, acting in opposite direction to the
relative motion in $x$ coordinates, is the generalized nonlinear
function
\begin{equation}\label{eq1}
F=f(\dot{x},z,t).
\end{equation}

The velocity argument can be seen as capturing the steady-state
friction behavior including the amplitude-constant and
$\mathrm{sign}(\dot{x})$-dependent Coulomb friction, the viscous
velocity-dependent friction, and Strtibeck velocity-weakening
curves as well. All three can be described by the well-known
steady-state characteristic curve
\begin{equation}\label{eq2}
F_{ss}(\dot{x})=\mathrm{sign}(\dot{x})\Bigl(F_{c}+(F_{s}-F_{c})
\exp\bigl(-|\dot{x}|^{\delta} V^{-\delta} \bigr)\Bigr) +\delta
\dot{x},
\end{equation}
often referred as static Stribeck friction model. The free
parameters are the Coulomb friction coefficient $F_{c}>0$,
Stribeck (or stiction) friction level $F_{s} > F_{c}$, linear
viscous friction coefficient $\delta \geq 0$, and two shape
factors of the velocity-weakening curve $V > 0$ and $\delta \neq
0$. For more details on the Stribeck effect and steady-state
characteristic friction curve (\ref{eq2}) we refer to
\cite{Stribeck1902,armstrong1994survey}.

The time-dependency of friction (\ref{eq1}) summarizes the weakly
known and often non-deterministic fluctuations in the frictional
behavior due to e.g. wear, adhesion effects, contact surface
irregularities, lubrication conditions, dust and others. Such
effects may cause some non-systematic parameters drifting of
friction modeling and, in what follows, are excluded from an
explicit consideration.

\begin{figure}
\centering
\includegraphics[width=0.65\columnwidth]{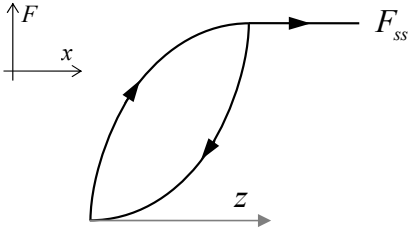}
\caption{Friction-displacement curve with hysteresis loop.}
\label{fig1}
\end{figure}

The $z$-argument represents an internal presliding state or
so-called relative presliding distance on the frictional
interface. Most simple way, this is a relative displacement at
each motion onset or motion reversal until the dynamic friction
force converges to the steady-state of gross sliding at an
unidirectional motion. Obviously, the presliding distance $z$ is
initialized (or reset) whenever the velocity sign changes and maps
explicitly the instantaneous state of presliding friction force
transitions. Depending on the particular form of presliding
friction $F=f(z)$, as a function of relative presliding distance,
an additional scaling factor $s$ may be used so that
\begin{equation}\label{eq3}
z=s \int \limits^{t}_{t_{r}} \dot{x} \, dt.
\end{equation}
Obviously, $t_{r}$ denotes the time instant of the last motion
reversal so that $|z|$ represents always a scaled distance to the
position where the motion direction changed for the last time. For
the sake of simplicity we will assume in the following $s=1$. The
friction-displacement curve of a contact surface at motion
reversals exhibit a hysteresis loop, see Figure \ref{fig1}.
The shape of hysteresis loops depends on multiple factors of the
asperities interaction, their elastic and plastic deformation and,
as a consequence, on energy dissipated on the frictional surface
during the motion cycles \cite{Ruder16a,ruderman2015presliding}.
According to \cite{koizumi1984study} the area of hysteresis loop
increases in proportion to the n-th power of presliding distance.
In particular, it has been found and experimentally proved that
this is the second power, i.e. $n=2$. Therefore, the curvature of
friction-displacement map during presliding is given by
\begin{equation}\label{eq4}
f(z)=z\bigl(1-\ln(z)\bigr).
\end{equation}
For details on deriving of equation (\ref{eq4}) from the above
$n$-th, respectably second, power condition we refer to
\cite{koizumi1984study}.

Assuming the hysteresis loop shape (\ref{eq4}) and $s=1$ it is
obvious that for zero initial state $F_{0}=0$ at the motion onset
the presliding friction is given by
\begin{equation}\label{eq5}
F(\dot{x},z)=\mathrm{sign}(\dot{x}) F_{ss}(\dot{x}) \,
z\bigl(1-\ln(z)\bigr).
\end{equation}
Assuming the presliding transitions always converge to the
steady-state $F_{ss}$ and the instantaneous friction value at the
last motion reversal is $F_{r}$ the friction force in presliding
is given by
\begin{equation}\label{eq6}
F(\dot{x},z)= \bigl| \mathrm{sign}(\dot{x}) F_{ss}(\dot{x}) -F_{r}
\bigr| \, z\bigl(1-\ln(z)\bigr)+F_{r}.
\end{equation}
Note that the normalized, through the scaling factor $s$,
presliding distance is defined on the interval $[-1,\,1]$, while
at the boundaries the friction force converges to the steady-state
value.

\section{Break-away conditions}
\label{sec:3}

The problem of break-away friction force can be seen as a problem
of detection (alternatively prediction) of the minimal actuation
force at which the motion system, being initially in the idle
state, begins the continuous (macro) motion, often denoted as
gross sliding. This problem is closely related to the stiction and
adhesion effects on the complex frictional interfaces and, at the
same time, is of empirical observation nature and relevance in the
engineering practice. The transitions from the system sticking to
gross sliding at an unidirectional motion have been observed in
various actuated machines and mechanisms and reported in e.g.
\cite{heslot1994creep,armstrong1994survey, lee2007static}. In most
the previously published works the varying break-away force (or
torque) has been exposed in dependency of the actuation (input)
force rate. That means the actuation (input) force has been
linearly increased starting from zero, i.e. $u=kt$, and the
break-away transition has been observed and recorded as when a
non-fluctuating quasi-constant acceleration occurs and the
relative velocity grows continuously. Note that before this, the
system is in presliding regime where a low relative displacement
can be detected, while the measured relative velocity is mostly
high-frequent oscillating around zero, with a relatively low
average and its increase, see e.g. experiments depicted in Fig. 8
in \cite{Rud10}.
\begin{figure}
\centering
\includegraphics[width=0.55\columnwidth]{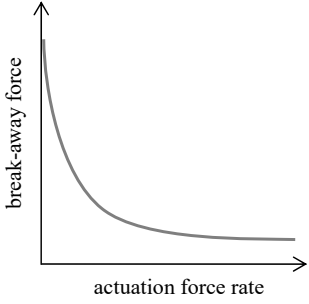}
\caption{Break-away force as function of the actuation force
rate.} \label{fig2}
\end{figure}
The dependency of the observed break-away force on the actuation
force rate $du/dt=k$ has been investigated and experimentally
demonstrated in
\cite{johannes1973role,iwasaki1999disturbance,lampaert2004experimental},
and also shown for the numerically simulated dynamic friction in
\cite{deWit1995new,iwasaki1999disturbance,swevers2000integrated}.
In all cases a typical inverse exponential map has been
highlighted as schematically shown in Figure \ref{fig2} (cf. e.g.
Fig. 4 in \cite{deWit1995new}, Fig. 5 in
\cite{iwasaki1999disturbance}, Fig. 13 in
\cite{swevers2000integrated}, Fig. 10 in
\cite{lampaert2004experimental}).

Now, we are in the position to analyze and describe analytically
the break-away conditions based on the modeling assumptions made
in Section \ref{sec:2}. We note that despite the break-away
dependency on the actuation force rate has been known from
experiments and confirmed by means of numerical simulations, no
explicit analytic form has been derived and validated so far in
line with the modeled presliding friction behavior.

For motion dynamics with a linearly increased actuation force we
write
\begin{equation}\label{eq7}
m\ddot{x}+f(\dot{x},z)=kt
\end{equation}
During presliding, the macroscopic system inertia can be neglected
due to a very low acceleration so that the actuation force is
mainly balanced by the counteracting friction, so that
$f(\dot{x},z)\approx kt$. Taking the time derivative of
(\ref{eq7}) and neglecting the inertial dynamics one obtains
\begin{equation}\label{eq8}
\frac{d}{dt}f(\dot{x},z)=k,
\end{equation}
while the full deferential yields \cite{Ruder2016b}
\begin{equation}\label{eq9}
\frac{d}{dt} f(\dot{x},z)=\frac{\partial f}{\partial
\dot{x}}\ddot{x} + \frac{\partial f}{\partial z}\dot{x}.
\end{equation}
For the same reason as above and due to the fact that $\partial
f/\partial \dot{x}=0$ within presliding, the first right-hand-side
summand in (\ref{eq9}) can be neglected and we obtain
\begin{equation}\label{eq10}
\frac{\partial f}{\partial z}\dot{x}-k=0.
\end{equation}
Substituting the derivative of (\ref{eq4}), with respect to $z$,
into (\ref{eq10}) results in
\begin{equation}\label{eq11}
-F_{ss} \ln(z)\dot{x}=k.
\end{equation}
It is obvious that the relative velocity during presliding
\begin{equation}\label{eq12}
\dot{x}=-\frac{k}{F_{ss}\ln(z)}
\end{equation}
can be computed as a function of relative presliding distance and
depends mainly on two factors $k$ and $F_{ss}$. While $k$ is fixed
for the given slope of external actuation force, the steady-state
friction value self depends on the instantaneous relative
velocity. Nevertheless, from (\ref{eq2}) we know that $F_{c} \leq
|F_{ss}| \leq F_{s}$ so that either both boundary values or an
average
$$
\hat{F}_{ss}=F_{c}+(F_{s}-F_{c})/2
$$
can be assumed for calculating (\ref{eq12}).

\begin{figure}
\centering
\includegraphics[width=0.98\columnwidth]{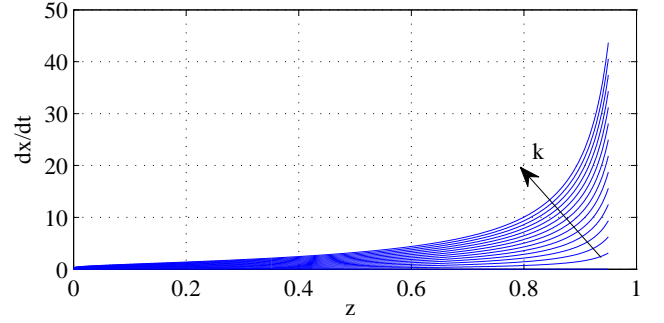}
\caption{Relative velocity as a function of presliding distance in
dependency of the rate of actuation force.} \label{fig3}
\end{figure}
The $(z,\dot{x})$ phase diagrams are shown in Figure \ref{fig3}
for the different actuation force rates $k$ and a fixed
$\hat{F}_{ss}$ value. One can see that for all actuation force
rates the relative velocity, starting from zero after a motion
reversal ($z=0$), increases exponentially when approaching
boundary of the presliding range ($z=1$). The computed phase
diagrams are for the actuation force rates $k \in
[0.01,\ldots,30]$ with an increment equal 2. Since a rapid
(exponential) increase of the relative velocity is for all $k$
when $z \rightarrow 1$ one can restrict the considered values by
e.g. 95 \% of presliding distance, denoted by $z_{0.95}$. Here we
should note that transition from the presliding to the gross
sliding is not abrupt/stepwise at all, and the break-away
conditions can be considered only for a certain, though
well-specified interval, like for example $0.95 < z < 1$ we
assumed. This is also in accord with the experimental and
numerical observations reported so far, while the break-away
detection is mostly realized ``at the time where a sharp increase
in the velocity could be observed'' \cite{deWit1995new}.

\begin{figure}
\centering
\includegraphics[width=0.98\columnwidth]{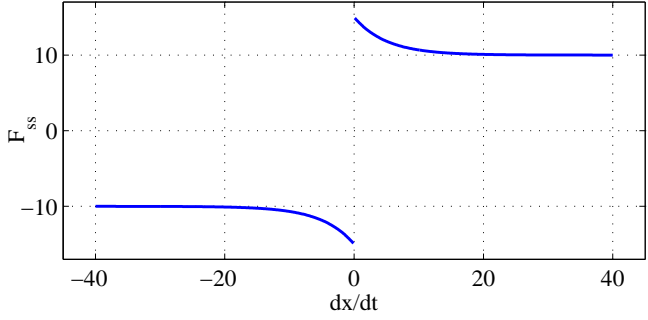}
\caption{Steady-state (Stribeck) characteristic friction curve.}
\label{fig4}
\end{figure}

\begin{figure}
\centering
\includegraphics[width=0.98\columnwidth]{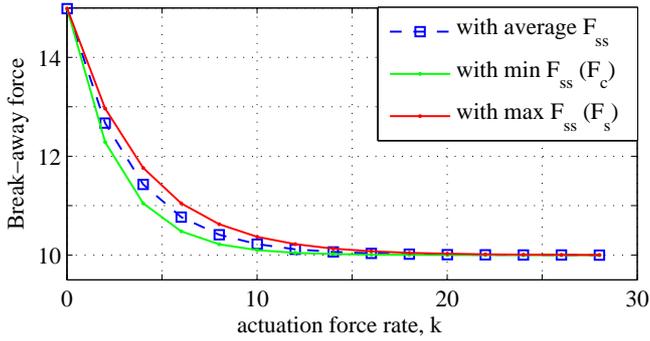}
\caption{Break-away force as function of the actuation force rate
computed by (\ref{eq2}), (\ref{eq12}), and (\ref{eq13}) for
minimal ($F_c$), maximal ($F_s$), and average ($\hat{F}_{ss}$)
steady-state friction values.} \label{fig5}
\end{figure}

For the assumed presliding boundary the break-away force can be
computed, based on (\ref{eq2}) and (\ref{eq12}), as
\begin{equation}\label{eq13}
F_{ba}=F_{ss}\bigl(\dot{x}(z_{0.95})\bigr).
\end{equation}
Now one can calculate the break-away force as a function of
actuation force rate $k$ and that provided the friction model
(\ref{eq2})-(\ref{eq4}) only is given. The assumed steady-state
(Stribeck) characteristic curve is depicted in Figure \ref{fig4}.
Note that the linear viscous friction coefficient $\sigma =0$ is
assumed for the sake of simplicity, and a relatively high
difference $F_{s} = 1.5 F_{c}$ between the minimal and maximal
steady-state friction values is chosen. The computed break-away
force as a function of actuation force rate is depicted in Figure
\ref{fig5}.
In order to reveal the impact of $F_{ss}$, assumed for
(\ref{eq12}) computation, the minimal ($F_{c}$), maximal
($F_{s}$), and averaged ($\hat{F}_{ss}$) steady-state values are
demonstrated opposite to each other. One can see that the
functional dependency of break-away force from $k$ is similar for
all three steady-state values. In particular, at lower (near to
zero) and higher actuation force rates their nearly coincide with
each other. On the opposite, the differences mostly occur at the
well pronounced exponential decrease.

\section{Conclusions}
\label{sec:4}

This notice aimed to analyze and describe analytically the
break-away conditions for which, at certain level of the external
actuation force and its rate, the presliding friction behavior
transits to the gross sliding and a continous (macro) motion sets
on. Using the straightforward formulation of the presliding and
steady-state friction force it has been explicitly shown how the
relative velocity progresses with the relative presliding distance
starting from zero idle state, and when a rapid exponential
increase of the relative velocity which is characteristic for
break-away occurs. We have derived an analytic expression for
computing the break-away force as a function of actuation force
rate. The computed and exposed results are in accord with the
previously published experimental observations and those from the
numerical simulations for which, however, an analytic expression
and analysis have been missed.

\bibliographystyle{elsarticle-num}        

\bibliography{references}

\begin{thebibliography}{10}
\expandafter\ifx\csname url\endcsname\relax
  \def\url#1{\texttt{#1}}\fi
\expandafter\ifx\csname urlprefix\endcsname\relax\def\urlprefix{URL }\fi
\expandafter\ifx\csname href\endcsname\relax
  \def\href#1#2{#2} \def\path#1{#1}\fi

\bibitem{altpeter1999friction}
F.~Altpeter, Friction modeling, identification and compensation, {PhD Thesis},
  EPFL (1999).

\bibitem{Dahl68}
P.~R. Dahl, A solid friction model, TOR 158(3107-18), The Aerospace
  Corporation, El Segundo (1968).

\bibitem{dahl1976solid}
P.~R. Dahl, Solid friction damping of mechanical vibrations, AIAA Journal
  14~(12) (1976) 1675--1682.

\bibitem{armstrong1994survey}
B.~Armstrong-H{\'e}louvry, P.~Dupont, C.~C. De~Wit, A survey of models,
  analysis tools and compensation methods for the control of machines with
  friction, Automatica 30~(7) (1994) 1083--1138.

\bibitem{deWit1995new}
C.~C. De~Wit, H.~Olsson, K.~J. Astrom, P.~Lischinsky, A new model for control
  of systems with friction, IEEE Transactions on automatic control 40~(3)
  (1995) 419--425.

\bibitem{iwasaki1999disturbance}
M.~Iwasaki, T.~Shibata, N.~Matsui, Disturbance-observer-based nonlinear
  friction compensation in table drive system, IEEE/ASME transactions on
  mechatronics 4~(1) (1999) 3--8.

\bibitem{swevers2000integrated}
J.~Swevers, F.~Al-Bender, C.~G. Ganseman, T.~Projogo, An integrated friction
  model structure with improved presliding behavior for accurate friction
  compensation, IEEE Transactions on automatic control 45~(4) (2000) 675--686.

\bibitem{johannes1973role}
V.~Johannes, M.~Green, C.~Brockley, The role of the rate of application of the
  tangential force in determining the static friction coefficient, Wear 24~(3)
  (1973) 381--385.

\bibitem{lampaert2004experimental}
V.~Lampaert, F.~Al-Bender, J.~Swevers, Experimental characterization of dry
  friction at low velocities on a developed tribometer setup for macroscopic
  measurements, Tribology Letters 16~(1-2) (2004) 95--105.

\bibitem{heslot1994creep}
F.~Heslot, T.~Baumberger, B.~Perrin, B.~Caroli, C.~Caroli, Creep, stick-slip,
  and dry-friction dynamics: Experiments and a heuristic model, Physical review
  E 49~(6) (1994) 4973.

\bibitem{socoliuc2004transition}
A.~Socoliuc, R.~Bennewitz, E.~Gnecco, E.~Meyer, Transition from stick-slip to
  continuous sliding in atomic friction: entering a new regime of ultralow
  friction, Physical review letters 92~(13) (2004) 134301.

\bibitem{lee2007static}
C.-H. Lee, A.~A. Polycarpou, Static friction experiments and verification of an
  improved elastic-plastic model including roughness effects, Journal of
  Tribology 129~(4) (2007) 754--760.

\bibitem{harnoy2008modeling}
A.~Harnoy, B.~Friedland, S.~Cohn, Modeling and measuring friction effects, IEEE
  Control Systems 28~(6) (2008) 82--91.

\bibitem{armstrong1993stick}
B.~Armstrong-Helouvry, Stick slip and control in low-speed motion, IEEE
  Transactions on Automatic Control 38~(10) (1993) 1483--1496.

\bibitem{Stribeck1902}
R.~Stribeck, Die wesentlichen {Eigenschaften} der {Gleit-} und {Rollenlager},
  VDI-Zeitschrift (in German) 46~(36--38) (1902)
  1341--1348,1432--1438,1463--1470.

\bibitem{Ruder16a}
M.~Ruderman, M.~Iwasaki, On damping characteristics of frictional hysteresis in
  pre-sliding range, Journal of Physics: Conference Series 727~(1) (2016)
  012014.

\bibitem{ruderman2015presliding}
M.~Ruderman, Presliding hysteresis damping of {LuGre} and {Maxwell-slip}
  friction models, Mechatronics 30 (2015) 225--230.

\bibitem{koizumi1984study}
T.~Koizumi, H.~Shibazaki, A study of the relationships governing starting
  rolling friction, Wear 93~(3) (1984) 281--290.

\bibitem{Rud10}
M.~Ruderman, T.~Bertram, Friction model with elasto-plasticity for advanced
  control applications, in: IEEE/ASME International Conference on Advanced
  Intelligent Mechatronics (AIM2010), 2010, pp. 914--919.

\bibitem{Ruder2016b}
M.~Ruderman, M.~Iwasaki, Analysis of linear feedback position control in
  presence of presliding friction, IEEJ Journal of Industry Applications 5~(2)
  (2016) 61--68.

\end{thebibliography}

\end{document}